\documentclass[%
reprint,
superscriptaddress,
floatfix,
amsmath,amssymb,
aps,
prx,
longbibliography
]{revtex4-2}
\usepackage{svg}
\usepackage{graphicx}% Include figure files
\usepackage{dcolumn}% Align table columns on decimal point
\usepackage{bm}% bold math
\usepackage{placeins}
\usepackage{float}
\usepackage{hyperref}% add hypertext capabilities
\usepackage{amsmath}
\usepackage{placeins}
\usepackage{times}
\usepackage{physics}
\usepackage{mathtools}
\usepackage{xcolor}
\usepackage{float}
\usepackage{hyperref}
\hypersetup{
    colorlinks,
    citecolor=blue,
    filecolor=black,
    linkcolor=blue,
    urlcolor=black
}
\usepackage[title,toc]{appendix}
\begin{document} 

\title{Versatile Full-Field Optical Coherence Tomography with Adjustable Transmission-to-Reflection Ratio and Enhanced Signal-to-Noise Ratio}

\author
{\normalsize{Youlong Fan,$^{1,\dag}$ 
Qingye Hu,$^{1,\dag}$
Zhongping Wang,$^{1}$ 
Zengming Zhang,$^{1}$ 
and Xiantao Wei$^{1\ast}$}\\
\small{$^{1}$School of Physical Sciences, University of Science and Technology of China, Hefei, 230041, China}\\
%\small{$^{2}$Department of Physics, University of Chicago, Chicago, IL 60637, USA}\\
\small{$^\dag$These authors contributed equally to this work.} \\
\small{$^\ast$To whom correspondence should be addressed; E-mail:~wxt@ustc.edu.cn.}
}

\date{\today}

\begin{abstract}
Traditional full-field optical coherence tomography (FF-OCT) is effective for rapid cross-sectional imaging but often suffers from incoherent signals due to imbalanced light intensities between the sample and reference arms. While the high-throughput dark-field (HTDF) FF-OCT technique employs an asymmetric beamsplitter (BS) to achieve an asymmetric beam-splitting ratio and optimize the utilization of available light, the fixed beam-splitting ratio in the optical system limits HTDF FF-OCT to effectively measuring only specific types of samples with certain scattering intensities. To address this limitation, we propose a more versatile FF-OCT system with an adjustable transmission-to-reflection ratio. This system enables accurate measurement across a broader range of samples by optimizing the light source and finely tuning the polarization to achieve the ideal ratio for different materials. We also observed that both signal-to-noise ratio (SNR) and imaging depth are influenced by the beam-splitting ratio. By precisely adjusting the beam-splitting ratio, both SNR and imaging depth can be optimized to achieve their optimal values.
\end{abstract}

\maketitle
\section*{}

In 1991, the Fujimoto group at MIT introduced Optical Coherence Tomography (OCT)\cite{RN24}, specifically Time-Domain OCT (TD-OCT), utilizing time-domain detection with point detectors and mechanically scanned reference arms, akin to a Michelson interferometer. This system employed low-coherence broadband light sources, splitting the light into reference and sample beams. Interference between the backscattered light and the reference beam necessitated pre-phase modulation and post-demodulation to enhance signal-to-noise ratios.

In 1995, Fercher et al. from the University of Vienna advanced the technology with Spectral-Domain OCT (SD-OCT)\cite{RN20}, which used a spectrometer to capture interference spectral signals, distinguishing it from TD-OCT. By 1997, Chinn and Swanson introduced Swept-Source OCT\cite{RN22}, leveraging grating-tuned semiconductor lasers for two-dimensional imaging, improving system resolution and depth perception. The following year, Beaurepair et al. proposed Full-Field OCT (FF-OCT)\cite{RN27}, offering a direct two-dimensional view of samples through coherent gating without scanning, allowing for rapid imaging and observation of interference structures.

%Despite its advantages, traditional FF-OCT \cite{RN27}\cite{RN14}is susceptible to incoherent signals due to significant intensity differences between backscattered light from the sample and reflected light from the reference arm. To address this, the high-throughput dark-field (HTDF) FF-OCT\cite{RN1} technique was developed, utilizing an asymmetric beamsplitter (BS) to suppress specular reflections and optimize light utilization. However, the selection of the BS's transmission-to-reflection ratio is highly dependent on the sample characteristics, and the currently available beamsplitters lack a continuously adjustable ratio, limiting the (HTDF) FF-OCT to specialized applications.

Despite its advantages, traditional FF-OCT is susceptible to incoherent signals due to significant intensity differences between backscattered light from the sample and reflected light from the reference arm. To address this, the high-throughput dark-field (HTDF) FF-OCT technique\cite{RN1} was developed, utilizing an asymmetric beamsplitter (BS) to suppress specular reflections and optimize light utilization. However, the best transmission-to-reflection ratio strongly dependent on sample characteristics, and the beamsplitters with fixed transmission-to-reflection ratio limits (HTDF) FF-OCT to specialized applications.

To address issues above, we propose a more versatile FF-OCT system with an adjustable transmission-to-reflection ratio. This system enables accurate measurement across a broader range of samples by optimizing the light source and finely tuning the polarization to achieve the ideal ratio for different materials. 

\begin{figure}[H]
    \centering
    \includegraphics[width=\linewidth]{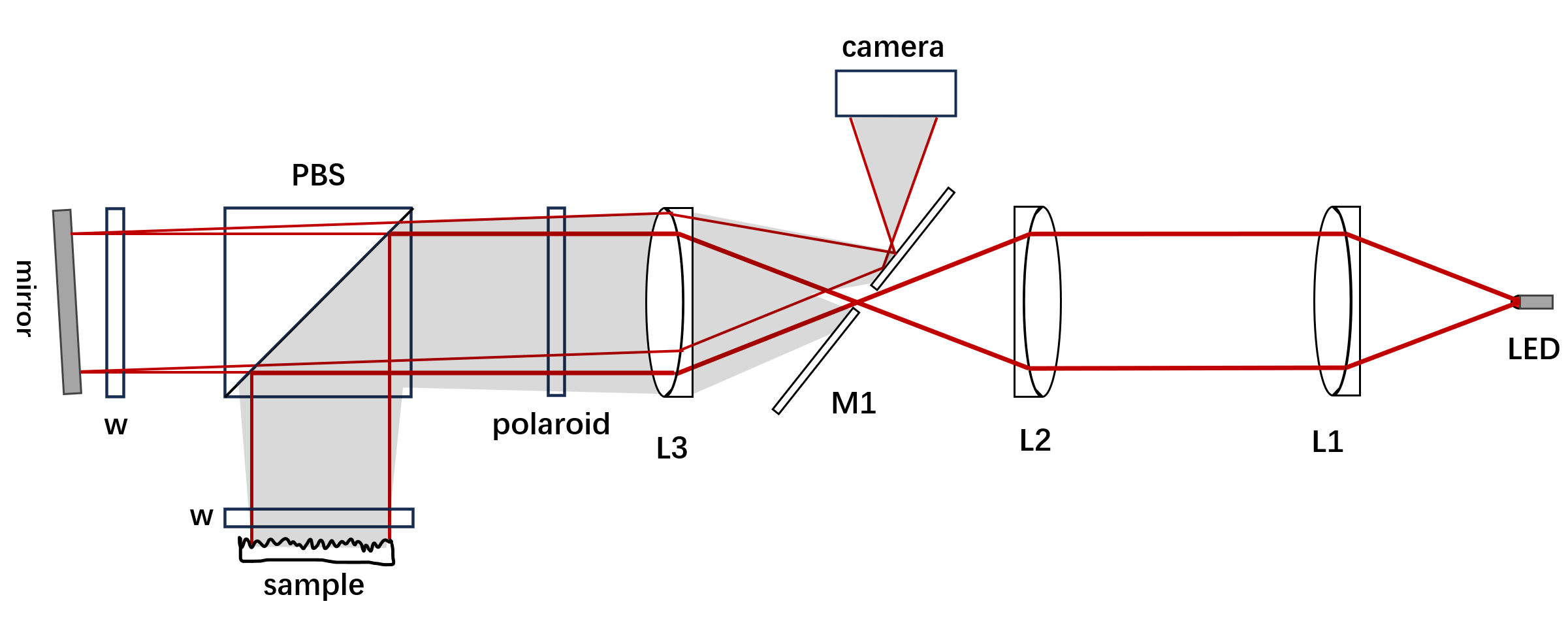}
    \caption{OCT setup. L1, doublet lens; L2, doublet lens; L3, doublet lens; M1, mirror with central aperture; w, glass window; Red beam shows illumination and the reference beam, whereas gray beams represent scattered light.}
    \label{fig1}
\end{figure}

Fig \ref{fig1} shows our experimental setup. The system utilizes LED\cite{RN34} (M730L5 and M850LP1, Thorlabs) as the light sources. M730L5 has a central wavelength of 730nm and a spectral bandwidth(FWHM) of 40 nm. M850LP1 has a central wavelength of 850nm and a spectral bandwidth(FWHM) of 30 nm. The emitted light from LED is focused to pass throuth a small hole located at the center of a mirror by doublet lenses L1 (f=10cm) and L2 (f=10cm). The mirror with a hole is positioned at the waist of the Gaussian beam after the reflected light from the reference arm passes through doublet lens L3 (f=10cm). A polarizing film is put between the lens L3 and one polarizing beamsplitter (PBS), converging mixed polarized light to linearly polarized light. The polarization direction forms an angle of $\theta$ with the vertical. After passing through the PBS, the vertical component of the light is reflected to the sample arm, while the horizontal component is transmitted to the reference arm. The polarization state of the backscattered light from the sample may vary, but the majority remains vertically polarized, which is reflected by the PBS and projected to the direction of polarizer. The light reflected from the reference arm is almost entirely horizontally polarized. Neglecting specular reflection and material absorption, it passes through the PBS with minimal loss and is projected in the same polarization direction as the polarizer. Due to a slight tilt in the reference arm's mirror, the light reflected back forms a Gaussian beam waist after passing through lens L3, bypassing the central aperture of mirror M1 and being reflected onto the a CMOS camera (MV-CA013-A0UM, Hikirobot), interfering with the backscattered sample light reaching the camera.
Next, a concise theoretical analysis will be conducted. Using a similar theoretical analysis as in HTDF FF-OCT\cite{RN1}, we can define the total number of photons emitted by the light source as \( N \). After passing through the polarizer, the number of photons becomes \( N/2 \). Subsequently, the number of photons allocated to the sample arm via the PBS is \( N/2 \cdot \cos^2\theta \), and the number of photons allocated to the reference arm is \( N/2 \cdot \sin^2\theta \) because of Malus law. In addition, we can define \( \cos^2\theta \) as \( r \) and \( \sin^2\theta \) as \( t \), so the ratio \( \sin^2\theta : \cos^2\theta = t : r \) represents the transmission-to-reflection ratio. Next, we define the coherent signal in the light backscattered from the sample as \( \frac{N}{2} \cdot \cos^2\theta \cdot R_O \). The incoherent signal is \( \frac{N}{2} \cdot \cos^2\theta \cdot R_{\text{inc}} \). The number of photons reflected back from the reference arm is \( \frac{N}{2} \cdot \sin^2\theta \cdot R_R \), and here \( R_R \) can be approximated as 1. Then after passing through polarizing film, both the light from reference and sample arm will be projected to the direction of polarizing film again. So the final photons reaching CMOS camera can be expressed as below:
%\begin{equation}
%\begin{aligned}
%N_{R} &=\frac{N}{2}\sin^{4}\theta R_{R},\\\\N_{O} &=\frac{N}{2}\cos^{4}\theta R_{O},\\\\N_{{\mathrm{inc}}} &=\frac{N}{2}\cos^{4}\theta R_{{\mathrm{inc}}},
%\end{aligned}
%\end{equation}
\begin{equation}
\begin{aligned}
N_{R} &=\frac{N}{2}\sin^{4}\theta R_{R},
\\\\N_{O} &=\frac{N}{2}\cos^{4}\theta R_{O},
\\\\N_{{\mathrm{inc}}} &=\frac{N}{2}\cos^{4}\theta R_{{\mathrm{inc}}},
\end{aligned}
\end{equation}
where \( N_R \) denotes the number of photons reflected back from the reference arm whereas \( N_O \) and \( N_{\text{inc}} \) represent the coherent and incoherent photon counts, respectively, originating from the sample arm. When we define SNR as
\begin{equation}\mathrm{SNR}\propto\frac{N_{R}N_{O}}{N_{R}+N_O+N_{{\mathrm{inc}}}+{nee}^2},\end{equation}
where $nee$ (noise equivalent electrons) refers to the effective electron count that quantifies the total electrical noise present in the CMOS camera. Since \( N_{\text{inc}} \) is relatively large, the terms \( nee \) and \( N_O \) in the denominator can be temporarily neglected. Next, substituting the specific expressions for \( N_R \), \( N_O \), and \( N_{\text{inc}} \) from equation (1) and supposing $R_R=1$, we obtain the SNR result for our OCT system as follows:
\begin{equation}
\mathrm{SNR}\propto\frac{N}{2}\frac{R_O\cos^4\theta\sin^4\theta}{\sin^4\theta+R_{\mathrm{inc}}\cos^4\theta}.
\label{eq:3}
\end{equation}
It can be observed that the SNR varies with \(\theta\), and the SNR reaches its maximum value when:
\begin{equation}
    \cos^2\theta=\frac{1}{1+\sqrt[3]{R_\text{inc}}}.
\end{equation}
Since the polarization direction \( \theta \) of the polarizer can be precisely adjusted, our system can, in theory, be tuned to the optimal \( \theta \) corresponding to the maximum SNR for different samples. This improves both the versatility and signal-to-noise ratio of the system.

Moreover, compared to traditional FF-OCT systems, our approach accounts for the effects of light polarization. After passing through the polarizer, the scattered light from the sample has the same polarization direction as the reflected light from the reference arm, filtering out the incoherent signals from the sample's backscattered light that have polarization directions different from the reference which cannot interfere. This enhances the signal-to-noise ratio. Additionally, the system's use of a mirror with a central aperture allows the surface reflections from optical components such as the cover glass of the sample, the PBS, and others to be focused by lens L3 and pass through the central aperture of mirror M1 without being detected by the CMOS camera. This effectively separates specular reflections from the signal to be collected, preventing the camera from overexposure, which allows the total photon count \( N \) in the experiment to be increased, thereby improving the overall signal-to-noise ratio.

To demonstrate the feasibility and superiority of our experimental system, we measured materials with distinct layered structures, such as adhesive tape, as well as living biological tissue, specifically a fingerprint. For each en-face picture of the cross-section, we moved the reflective mirror at the corresponding position on the nanometer scale to induce phase changes of $0^\circ$, $90^\circ$, $180^\circ$, and $270^\circ$, corresponding to intensity values $I_1$, $I_2$, $I_3$, and $I_4$, respectively. To eliminate the effects of vibration and other factors, thereby further improving the SNR, the intensity value \(I\) was obtained by continuously capturing four images at each phase and averaging them. The cross-section image was then obtained by calculating the equation $\sqrt{\left(I_1-I_3\right)^2+\left(I_2-I_4\right)^2}$\cite{RN7}. Then, the stepping motorized translation stage can be moved over a wide range to adjust the sample arm and reference arm to nearly equal optical paths, determining the region where coherent signals are present. Afterward, the jog distance of each step of the motor is reduced to increase resolution, allowing for a finer scan within the coherent signal region to obtain cross-sectional images of the sample at different depths.

%In this experiment, the displacement step of the motor-driven stage was initially set to 20 micrometers, covering a relatively large total displacement range. At each position within this range, images were captured at four distinct phase intervals, with four images taken for each phase. The obtained images were then processed and analyzed using four-phase subtraction to identify the displacement range generating coherent signals. Based on this analysis, the step size of the motor-controlled displacement stage was reduced to 6 micrometers, and the total displacement range was confined to the region producing coherent signals. The acquired images were subsequently averaged to reduce noise, and phase subtraction was applied to the four phases at each position to generate two-dimensional cross-sectional images.

\begin{figure}[H]
    \centering
    \includegraphics[width=\linewidth]{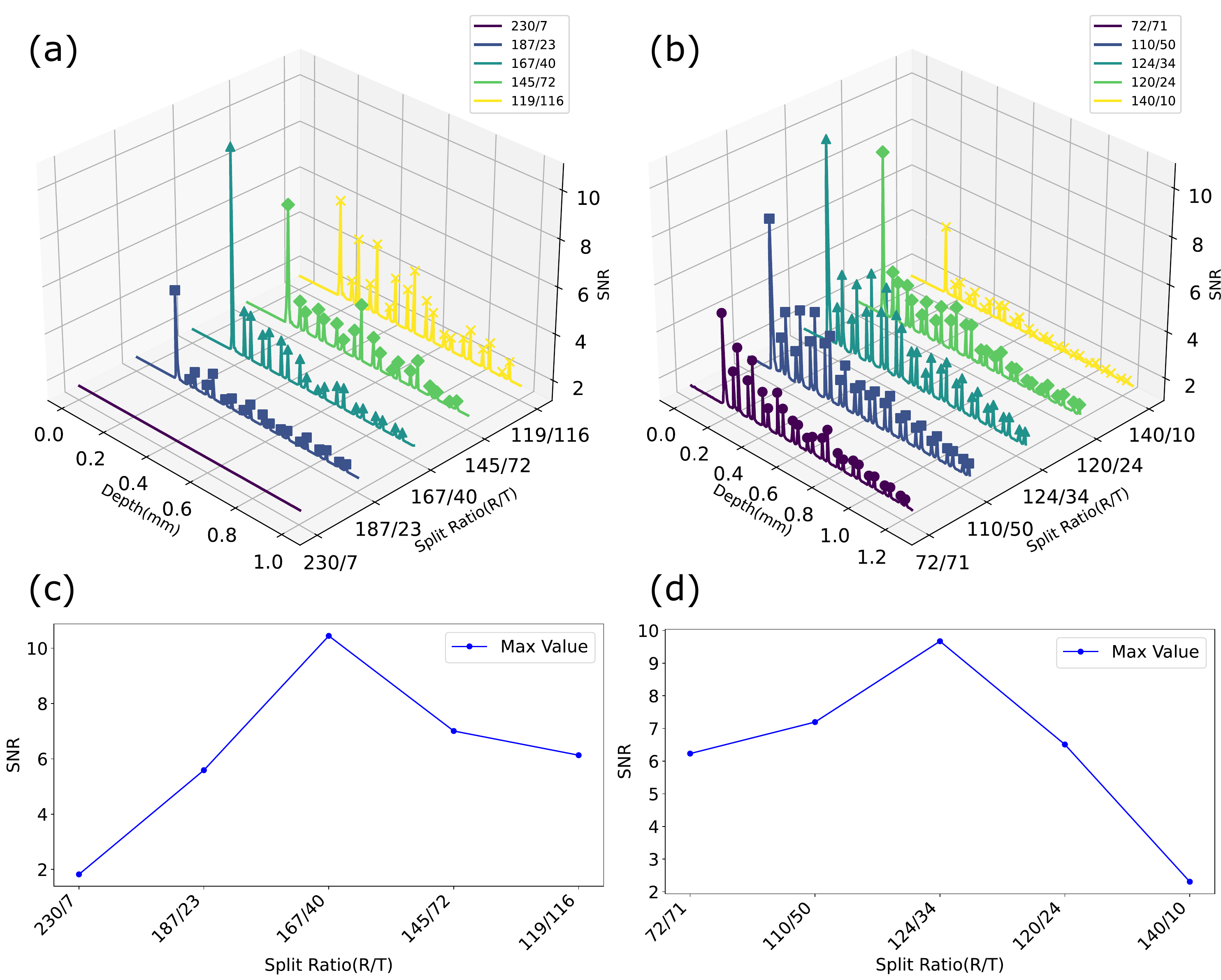}
    \caption{SNR versus depth images (a) and (b) acquired from the same tape using PBS-split OCT. (a) Images obtained by varying the splitting ratio while keeping the light source intensity constant. (b) Images obtained by varying the splitting ratio while keeping the signal intensity detected by CMOS camera constant. The corresponding relationship between the tape surface signal and the splitting ratio is shown in (c) and (d). (c) Images generated by varying the splitting ratio with constant light source intensity. (d) Images generated by varying the splitting ratio with constant signal intensity detected by CMOS camera.}
    \label{fig:tape}
\end{figure}

Furthermore, by varying the polarization direction of the polarizer, different splitting ratios were achieved by the combination of polarizing film and PBS. Thus, we can obtain information about the sample at different depths and with varying beam-splitting ratios, as shown in Fig\ref{fig:tape}(a) and Fig\ref{fig:tape}(b). The SNR at the same depth was extracted to establish the correlation between the SNR and the PBS splitting ratio, as shown in Fig\ref{fig:tape}(c) and Fig\ref{fig:tape}(d).

Regarding the method of representing SNR in the experiment, we extracted the region with coherent signals from the cross-sectional image for every selected beam-splitting ratio. The mean value of the pixel intensities in this region was calculated and used as the numerator for SNR. For the denominator, cross-sectional images were captured from a depth without back scattered light(where $R_o=0$), and the region identical to that selected for the coherent signals was extracted. The standard deviation of the pixel intensities in this region calculated, and  used as the denominator for the SNR.

%The expression for the signal-to-noise ratio (SNR) in the experiment is as follows: We extracted the coherent signal region S from the cross-sectional image for each selected splitting ratio and calculated the mean pixel intensity A of region S. From a non-coherent location, we captured the cross-sectional image and extracted the same region S corresponding to the coherent signal. We calculated the mean pixel intensity for this region and then obtained the standard deviation B of these means. We define SNR as SNR = A/B.%?

Fig\ref{fig:tape}(a) with the method of maintaining constant light source intensity clearly demonstrates the dependence of SNR on the beam-splitting ratio, as shown in Equation \ref{eq:3}. In Fig. \ref{fig:tape}(b), controlling the received light intensity to remain non-saturated, relatively high, and constant is intended to demonstrate the optimal SNR achievable at each beam-splitting ratio. This is because, as shown in Equation \ref{eq:3}, SNR is proportional to the total photon count \(N\).

%In Fig\ref{fig:tape}(a)(b), the layered structure of the adhesive tape can be observed. Fig\ref{fig:tapepi} presents two-dimensional side view images of two different types of high-temperature-resistant insulating tape used as samples. A single layer of tape is composed of a substrate and adhesive, with a strong scattering signal at the interface between the substrate and the adhesive, corresponding to a higher SNR. In contrast, within the relatively uniform substrate or adhesive, only weak scattering signals caused by material inhomogeneities or defects are present. These coherent signals are relatively faint and are mostly submerged in background noise. It can also be observed that as the penetration depth increases, the SNR gradually decreases. However, since the tape is a material with good transparency, the loss during propagation within the same layer is minimal, allowing for a penetration depth close to the centimeter scale. At the same time, it is also evident that the beam-splitting ratio has a significant impact on both the SNR and the penetration depth. Based on Fig\ref{fig:tape}(c)(d), it can be concluded that the optimal R:T for this tape is around 4:1.原文

\begin{figure}[H]
    \centering
    \includegraphics[width=\linewidth]{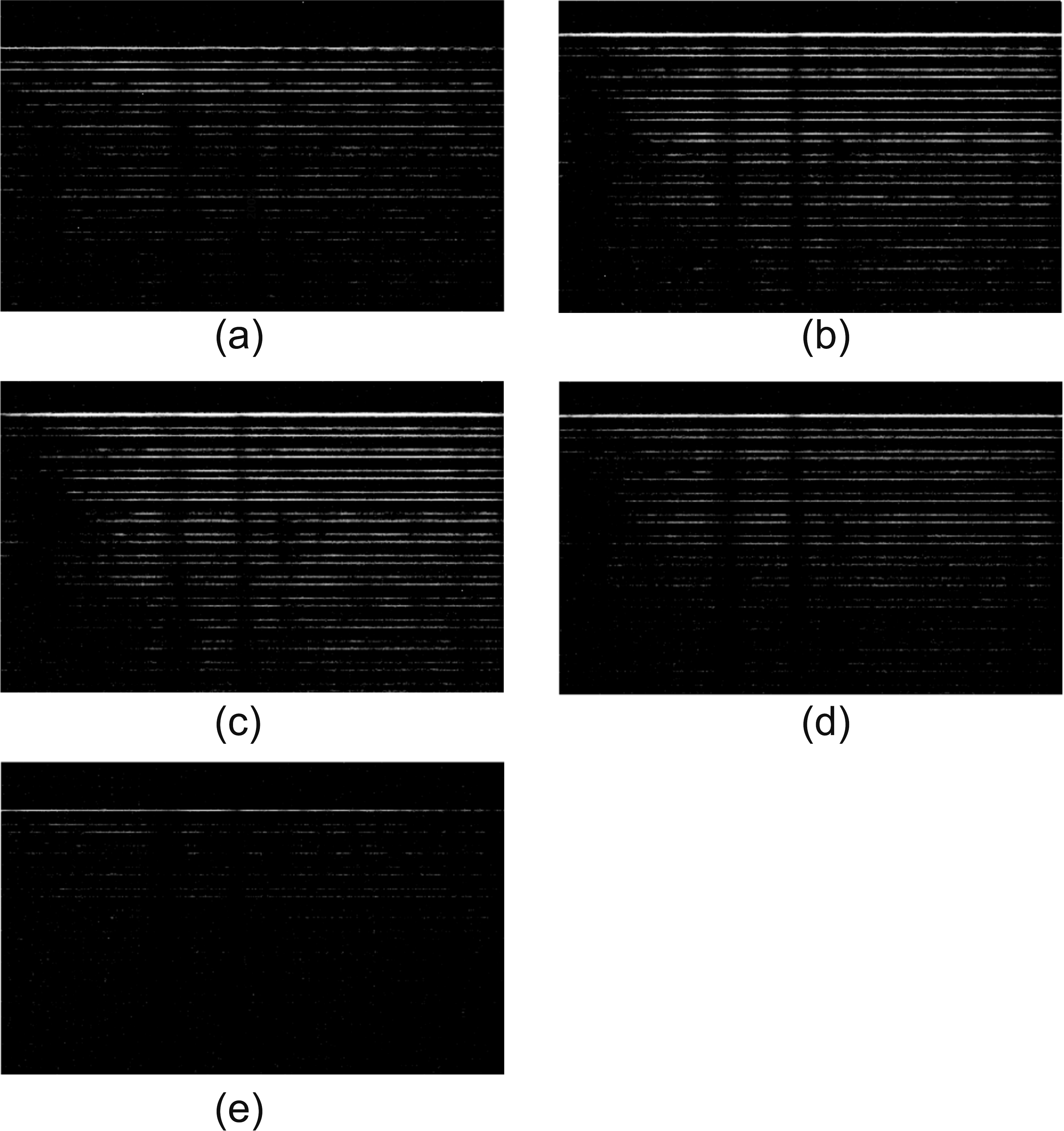}
    \caption{Image(a)-(e) are longitudinal cross-sectional images of 80mm thick tape under different beam-splitting ratios, where (a) corresponds to a beam-splitting ratio of 72/71, (b) corresponds to a ratio of 110/50, (c) corresponds to a ratio of 124/34, (d) corresponds to a ratio of 120/24, and (e) corresponds to a ratio of 140/10.}
    \label{fig:5tape}
\end{figure}

In Fig\ref{fig:tape}(a)(b), the layered structure of the adhesive tape can be observed. Fig\ref{fig:5tape} presents two-dimensional side view images of high-temperature-resistant insulating tape used as samples. A single layer of tape is composed of a substrate and adhesive, with a strong scattering signal at the interface between the substrate and the adhesive, corresponding to a higher SNR. In contrast, within the relatively uniform substrate or adhesive, only weak scattering signals caused by material inhomogeneities or defects are present. These coherent signals are relatively faint and are mostly submerged in background noise. It can also be observed that as the penetration depth increases, the SNR gradually decreases. However, since the tape is a material with good transparency, the loss during propagation within the same layer is minimal, allowing for a penetration depth close to the centimeter scale. At the same time, it is also evident that the beam-splitting ratio has a significant impact on both the SNR and the penetration depth. Based on Fig\ref{fig:tape}(c)(d), it can be concluded that the optimal R:T for this tape is around 4:1.

%\begin{figure}[H]
   % \centering
   % \includesvg[width=\linewidth]{figs/tapepic.svg}
   % \caption{(a) Tape with a total thickness of 80 $\mu m$, with a substrate thickness to adhesive thickness ratio of 5:3.
   % (b) Tape with a total thickness of 100 $\mu m$, with a substrate thickness to adhesive thickness ratio of 3:1.}
    %\label{fig:tapepi}
%\end{figure}

\begin{figure}[H]
    \centering
    \includegraphics[width=\linewidth]{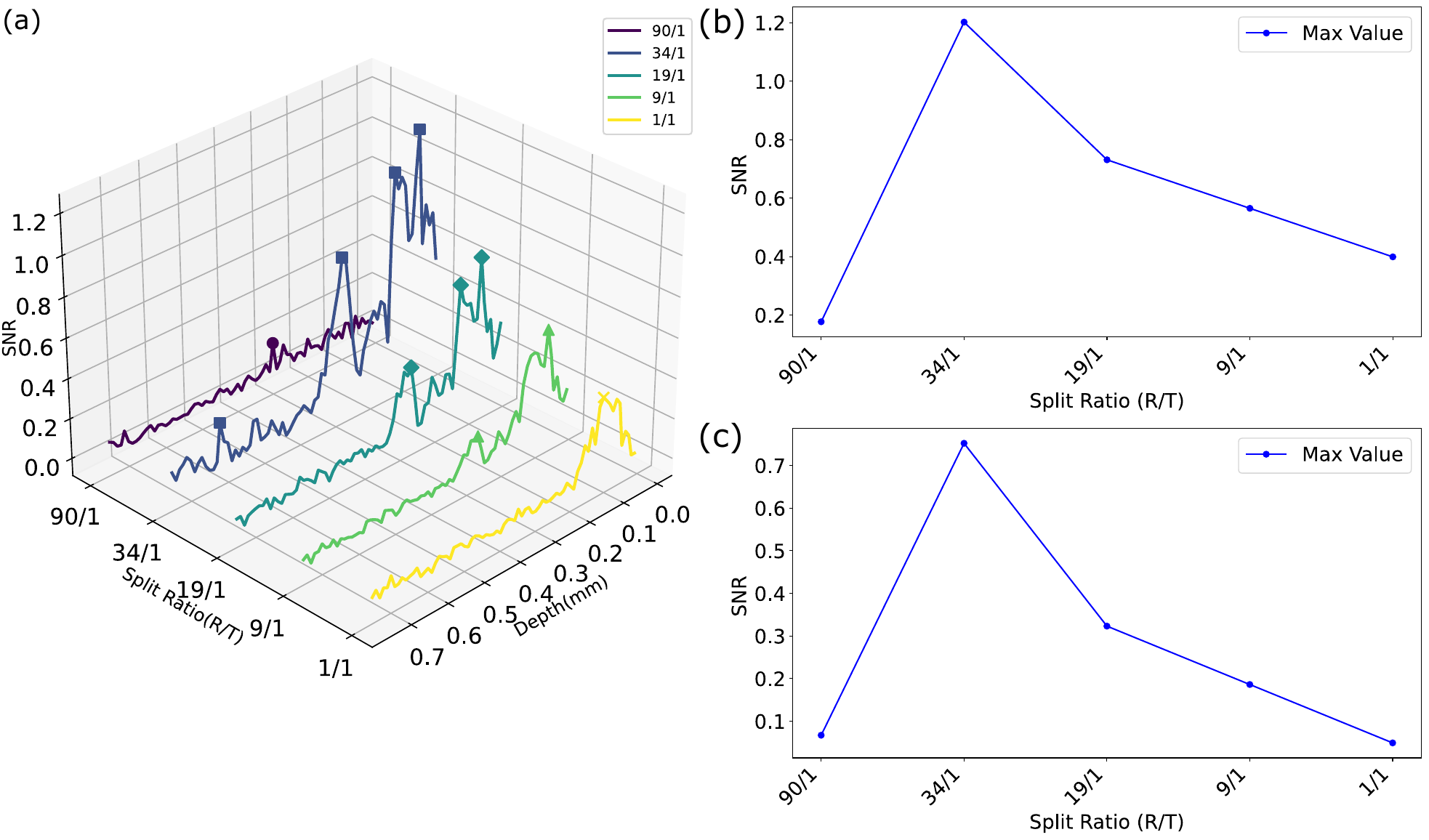}
    \caption{Image (a) shows the relationship between SNR and depth when the sample is a fingerprint, as well as the effect of varying the splitting ratio on this relationship. Image (b) displays the SNR values of external fingerprints under different splitting ratio conditions, while Image (c) shows the SNR values of internal fingerprints under different splitting ratio conditions.}
    \label{fig:fin}
\end{figure}

\begin{figure}[H]
    \centering
    \includegraphics[width=\linewidth]{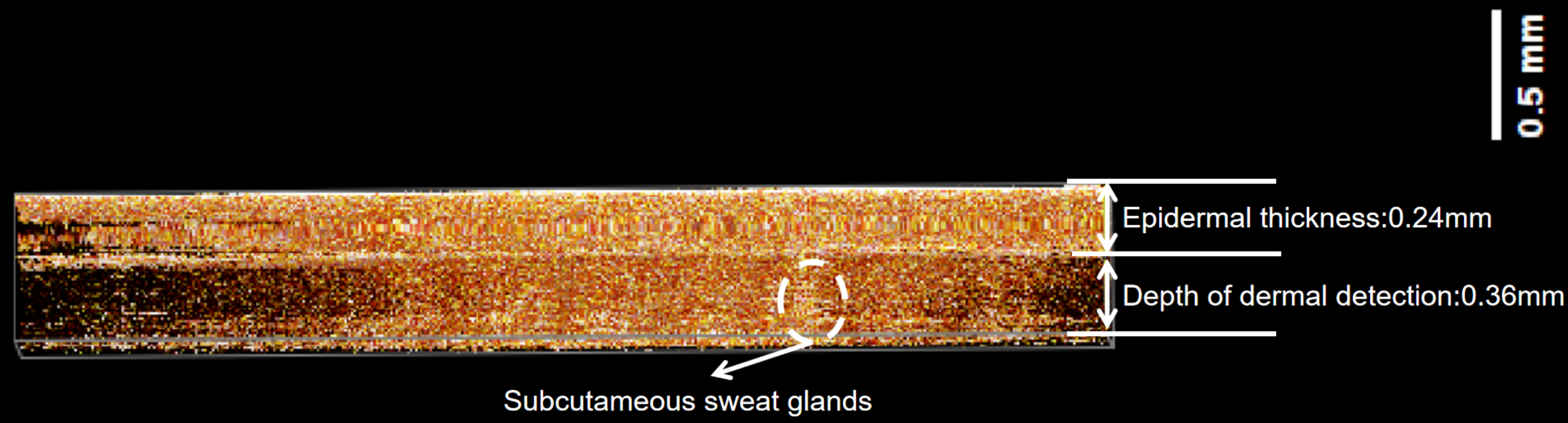}
    \caption{Side view of a fingerprint}
    \label{fig:sidefin}
\end{figure}

The experimental system was also used to measure the fingerprint\cite{RN2,RN54,RN21}, which is living biological tissue. Compared to the previous tape measurement, we simply replaced the light source from the M730L5 to the near-infrared source M850LP1, which has better penetration and is less absorbed by biological tissue. Additionally, since biological tissue has weaker and less directional backscattering, the R:T ratio needed to be increased compared to the tape measurement. Fig\ref{fig:fin}(a) shows the relationship between SNR and depth at different beam-splitting ratios. For fingerprint measurements, the optimal beam-splitting ratio is around 34:1. At this ratio, the dual-layer structure of the fingerprint, consisting of the epidermis and dermis, can be clearly distinguished from the waveform. When the beam-splitting ratio is 90:1, there is no coherent signal from the backscattering light of the fingerprint. At a ratio of 1:1, only the signal from the epidermis is present. Thus, the beam-splitting ratio affects both the SNR intensity and the imaging penetration depth. Fig\ref{fig:sidefin} shows the side view of the fingerprint, where the dual-layer structure of the epidermis and dermis can be observed. The epidermis is approximately 0.24 mm thick, and the effective detection depth of the dermis with this experimental system is 0.60 mm. The structure of the sweat glands within the dermis is also visible.
\vspace{-0.1cm}

Previously, a combination of a fixed asymmetric beamsplitter and a neutral density filter was employed to achieve a discontinuously adjustable R:T ratio. In comparison, the current system features a more streamlined and simplified optical setup. For tape measurements, the optimal SNR in the previous system was achieved at a R:T ratio of 3.15:1, whereas the new system delivers an SNR that is 1.17 times higher at a ratio of 124:34. For fingerprint measurements, the previous system reached optimal SNR at a ratio of 26:1, while the new system, with a ratio of 34:1, results in an SNR improvement of 1.03 times for the epidermis and 1.01 times for the dermis. Additionally, the imaging depth increased from 0.44 mm to 0.60 mm. 

In conclusion, a FF-OCT system with an adjustable T:R ratio was utilized to achieve improved SNR and enhanced imaging depth. The new versatile FF-OCT system achieves high-resolution adjustments of the R:T ratio simply by rotating the polarizer to control the polarization direction. By considering polarization effects in the interference process, the system ensures that light from both the sample and reference arms reaches the CMOS camera with the same polarization direction. This approach effectively filters out incoherent signals from light with mismatched polarization, combined with a central-aperture mirror that allows specular reflections from the surfaces of the PBS, lens, and other optical components to pass directly through the aperture without reaching the CMOS camera, improving SNR while preventing camera overexposure. As a result, the incident light intensity can be increased further, enabling greater penetration depth into the sample.

\section*{Acknowledgments}
This work was financially supported by National Natural Science Foundation of China (No. 62375255).

%Add bibliography here
%\bibliographystyle{unsrt}  
\bibliography{OCTv2}

\end{document}